\documentclass[12pt,A4,twoside]{article}
\usepackage[utf8]{inputenc}
\usepackage{amssymb}
\usepackage{amsmath}
\usepackage{amsfonts}
\usepackage{amsbsy}
\usepackage{natbib}
\usepackage{eucal}
\usepackage{graphicx}
\usepackage{titlesec}
\usepackage[english]{babel}
\usepackage{times}
\usepackage{titling}
\usepackage{mathptmx}
\usepackage{float}
\usepackage[font=small]{caption}
\usepackage{array}
\usepackage[colorlinks,citecolor=blue,urlcolor=blue]{hyperref}
\usepackage{authblk}
\usepackage{amsthm}

\usepackage{booktabs}
\usepackage{subcaption}

\usepackage{setspace}
\DeclareSymbolFont{txgreek}{OML}{cmr}{m}{it}

\renewcommand{\abstract}[1]{{\small\noindent
\hrulefill\par \vspace*{0.1cm}\noindent{\small\bf\sffamily
{Abstract}}\parindent=0pt\par\noindent\vspace{-0.1cm}\noindent\hrulefill\par\vspace*{0.5\baselineskip}\hspace*{0cm}\renewcommand{\baselinestretch}{1.1}\sffamily{#1}\par\vspace*{-0.1cm}\noindent\hrulefill}}

\def\and{,\;}
\DeclareMathSizes{11}{11}{7.8}{6.5}

\def\paragraf{\fontsize{9}{10pt}\fontfamily{phv}\fontshape{it}\selectfont}
\titleformat{\paragraph}
{\titlerule[0pt]
\vspace{0cm}%
\paragraf}
{\theparagraph}{.5em}{\vspace*{0\baselineskip}}

\def\titol{\fontsize{12.045}{12pt}\fontfamily{phv}\fontseries{b}\selectfont}
\titleformat{\section}
{\titlerule[0pt]
\vspace{0.2cm}%
\titol}
{\thesection.}{.5em}{\vspace*{0\baselineskip}}

\def\titolp{\fontsize{11.045}{11pt}\fontfamily{phv}\fontseries{b}\fontshape{it}\selectfont}
\titleformat{\subsection}
{\titlerule[0pt]\bigskip
\vspace{-0.4cm}%
\titolp}
{\thesubsection.}{.5em}{\vspace*{0\baselineskip}}

\def\titolpp{\fontsize{10.045}{10pt}\fontfamily{phv}\fontshape{it}\selectfont}
\titleformat{\subsubsection}
{\titlerule[0pt]
\vspace{-0.4cm}%
\titolpp}
{\thesubsubsection.}{.5em}{\vspace*{0\baselineskip}}

\usepackage{titling}

\pretitle{\begin{center}\sffamily\fontsize{18pt}{20pt}\selectfont}
    \posttitle{\par\end{center}}
    \preauthor{\begin{center}\fontsize{12pt}{14pt}\selectfont}
    \postauthor{\par\end{center}\vspace{0bp}}
    \predate{}
    \date{}
    \postdate{}

\title{Predicting subjective rage and facial expressions in human driving: A Bayesian network approach with beta-distributed nodes}

\thanksmarkseries{arabic}
\author{Zaïra Méndez-Porcar\thanks{Departament d'Estadística i Investigació Operativa, Universitat de València, València, Spain} \and  Francisco Palmí-Perales$^1$\and  Gabriel Calvo$^1$\and  Carmen Armero$^1$\and Ana de la Torre-García\thanks{Instituto de Biomecánica de Valencia, Universitat Politécnica de Valencia, Valencia, Spain}}

\def\headers#1{\fontsize{8.5}{10}\centering\sffamily\itshape{#1}}
\def\page#1{\fontsize{8.5}{10}\sffamily{#1}}
\usepackage{fancyhdr}

\begin{document}
\maketitle

\thispagestyle{empty}
\renewcommand{\headrulewidth}{0truecm}
\pagestyle{fancy}
\rhead[\headers{Predicting subjective rage and facial
expressions in human driving...}]{\page{\thepage}}
\lhead[\page{\thepage}]{\headers{Z. Méndez-Porcar et al.}}
 \lfoot{} \rfoot{}
\cfoot{}

\abstract{A Bayesian network framework is proposed for modelling unit-bounded continuous variables using conditional beta-distributed nodes within a fully Bayesian inference setting. The model captures conditional dependencies and propagates uncertainty through the network, with inference performed via Markov Chain Monte Carlo methods implemented in WinBUGS. The framework is applied to an experimental study of emotional and facial responses, focusing on rage intensity and facial gestures. Results show that brow lowering is strongly associated with rage intensity and is more frequent in men, whereas upper lid raising decreases under provocation independently of rage or sex. The model also predicts rage severity from informative facial gestures.}


\paragraph{Keywords: Bayesian inference and prediction; beta mixed regression models; Facial Acting Code
System (FACS); semi-autonomous driving}

\renewcommand{\baselinestretch}{1.2}
\bigskip

\section{Introduction}\label{sec1}
Bayesian networks (BNs), also referred to as belief networks, are a subclass of probabilistic graphical models that represent joint probability distributions of  random vectors \citep{pearl1995Bayesian}. The primary strength of BNs lies in their capacity to represent conditional dependency relationships among random variables within a coherent and quantitative uncertainty framework \citep{wang2023quantification}. The structure of a BN follows a directed acyclic graph (DAG), which facilitates the visualisation of causal or associative interactions, as well as the factorisation of the joint probability distribution of all the random variables involved into local conditional distributions. This property provides relevant computational advantages by simplifying inference calculations and allowing efficient updates when new observations are incorporated. In addition, this modular structure makes it possible to isolate and interpret the effect of individual variables or substructures within the network, which is particularly useful in complex systems \citep{cowell1999probabilistic}.

BNs have been successfully applied across a wide range of domains, including medicine, neuroscience, environmental sciences, and computational biology, where they are used for tasks such as prediction, diagnosis, decision support, and uncertainty quantification \citep{babakov2025reusability}. Despite the flexibility and applicability of BN, its practical implementation is often constrained by the capabilities of available software. For instance, packages such as \texttt{bnlearn} \citep{scutari2010learning} are widely used for structure learning and inference but are typically restricted to a limited set of node types, mainly discrete and Gaussian. Other frameworks, such as \texttt{abn} \citep{kratzer2023additive}, extend this functionality by allowing additional probability distributions. However, they still lack support for other commonly used distributions.

This paper deals with the use of Bayesian inference and prediction for BN statistical learning. We focus on their use for modelling unit bounded continuous data through the incorporation of beta mixed regression models.  While these models are highly relevant for applications where observations are naturally constrained to the unit interval, they remain relatively underexplored within the BN literature. In this context, we apply the proposed approach to an experiment from the European BERTHA project (\url{https://berthaproject.eu}), focusing on the modelling and prediction of some mental states of a person from their facial behaviour. The BERTHA project focuses on developing a scalable, probabilistic driver behavioural model to enable safer and more human-like decision-making in connected and automated vehicles. Ultimately, this work, within the BERTHA project, aims to infer a driver’s underlying mental state by analyzing their facial gestures and expressions within real-world driving contexts.

The main contribution of this paper is twofold. First, the extension of the standard BN framework by introducing beta-distributed nodes within a mixed regression setting. Although beta distributions have been used in BN-related contexts, particularly to represent uncertainty about probabilities or model parameters \citep{needham2007primer}, their use as response variables in continuous BN structures remains uncommon. Second, we formulate inference and prediction for this extended model within a fully Bayesian framework. This allows parameter uncertainty to be propagated coherently through the network, rather than relying only on point estimates.

  The remainder of this paper is organized as follows. Section 2 introduces the methodological framework, where Subsection 2.1 presents a general Bayesian network (BN) formulation for Bayesian learning, and Subsection 2.2 incorporates beta mixed regression models into the BN framework. Section 3 shifts focus to the emotional domain of human driving, specifically examining rage and facial expressions; within this section, Subsection 3.1 describes the experimental context, Subsection 3.2 details the gathered dataset, and Subsection 3.3 outlines the BN modelling strategy, evaluating two distinct proposals regarding the nature of a specific network variable. Section 4 evaluates and compares the predictive performance of these two modelling approaches, presenting the posterior results for both estimation and prediction. Finally, Section 5 outlines the conclusions.

\section{Methodological framework}\label{sec2}

\subsection{Bayesian inference for Bayesian networks}\label{subsec21}

BN are probabilistic graphical models designed to compactly describe dependency structures among a set of random variables. Each BN is represented by a DAG in which nodes correspond to random variables and edges encode conditional dependency relationships among them. This framework formalizes uncertainty and quantifies the causal or associative influence among variables through conditional probability distributions.

A BN for the set of random variables $\boldsymbol{Y} = \{ Y^{(v)}, v \in \mathcal{V}\}$,  where $\mathcal{V} = \{1, 2, \ldots, V\}$ denotes the set of nodes in the graph, random effects $\boldsymbol{\varphi}$ and parameters and hyperparameters $\boldsymbol{\theta}$ define a joint probability distribution over these three elements:

\begin{equation}
f(\boldsymbol{y}, \boldsymbol{\theta}, \boldsymbol{\varphi}) = f(\boldsymbol{y} \mid  \boldsymbol{\theta}, \boldsymbol{\varphi}) \, f(\boldsymbol{\varphi} \mid \boldsymbol{\theta}) \, \pi(\boldsymbol{\theta}),
\end{equation}

\noindent where $f(\boldsymbol{y} \mid \boldsymbol{\theta}, \boldsymbol{\varphi})$ is the conditional distribution of the observable nodes of the BN given $(\boldsymbol{\theta}, \boldsymbol{\varphi})$, $f(\boldsymbol{\varphi} \mid \boldsymbol{\theta})$ the conditional distribution of the random effects $\boldsymbol{\varphi}$ given $\boldsymbol{\theta}$ and $\pi(\boldsymbol{\theta})$ is the prior distribution of $\boldsymbol{\theta}$.

The graphical essence of a BN explicitly establishes a hierarchical structure among its random variables. If there exists a directed edge from node $Y^{(v)}$ to $Y^{(u)}$, then $Y^{(v)}$ is considered a \textit{parent} of $Y^{(u)}$, and conversely, $Y^{(u)}$ is a \textit{child} of $Y^{(v)}$. Based on these relationships, the concepts of \textit{ancestors}  and \textit{descendants}  of a node can be formally defined. These notions are essential for analyzing how information propagates throughout the network \citep{ben2008Bayesian}. It is worth noting that $\boldsymbol{\theta}$ and $\boldsymbol{\varphi}$ can also be parents and children in Bayesian learning, but for simplicity, we omit them.

The most convenient property of Bayesian networks is that they allow the factorisation of the joint conditional probability distribution $f(\boldsymbol{y} \mid \boldsymbol{\theta}, \boldsymbol{\varphi})$ in terms of local conditional distributions, exploiting the conditional independence relationships encoded in the DAG \citep{cowell1999probabilistic}. Thus, if $Pa(Y^{(v)})$ denotes the set of parents of variable $Y^{(v)}$, then:
$$
f(\boldsymbol{y} \mid \boldsymbol{\theta}, \boldsymbol{\varphi}) = \prod_{v \in \mathcal{V}} f(y^{(v)} \mid Pa(y^{(v)}),  \boldsymbol{\theta}, \boldsymbol{\varphi}).
$$

This factorisation implies conditional independence among the random variables of the BN given their parents, $\boldsymbol{\theta}$ and $\boldsymbol{\varphi}$. This property constitutes the basis for the computational efficiency of Bayesian networks, as it allows decomposing a complex problem into local submodels. 

Within the Bayesian statistical learning, inference is rigorously carried out through the computation of the posterior distribution of the model parameters, hyperparameters $\boldsymbol{\theta}$ and the vector $\boldsymbol{\varphi}$ of random effects. This approach combines prior information and empirical evidence within a unified probabilistic framework through Bayes' Theorem. 

Furthermore, BNs provide a natural environment for predicting new experimental outcomes, $Y_\star$, via joint posterior predictive distributions. This architecture facilitates the derivation of both conditional and marginal posterior predictive distributions, a capability that follows directly from the factorisation properties of the graphical model. In a Bayesian setting, prediction is formally obtained through the joint posterior predictive distribution defined as
\begin{equation*}
\begin{aligned}
f(\boldsymbol{y}_{\star} \mid \mathcal{D})
&=
\int f(\boldsymbol{y}_{\star} \mid \boldsymbol{\varphi}, \boldsymbol{\theta})
\, \pi (\boldsymbol{\varphi}, \boldsymbol{\theta} \mid \mathcal{D})
\, \text{d}(\boldsymbol{\varphi}, \boldsymbol{\theta}) \\
&=
\int \prod_{v\in\mathcal{V}}
f(y_{\star}^{(v)}\mid Pa(y_{\star}^{(v)}),\boldsymbol{\varphi},\boldsymbol{\theta})
\, \pi(\boldsymbol{\varphi},\boldsymbol{\theta} \mid \mathcal{D})
\, \text{d}(\boldsymbol{\varphi},\boldsymbol{\theta}).
\end{aligned}
\end{equation*}

This predictive distribution can be approximated through simulation using the sampling models, $f(  y_{\star}^{(v)} \mid Pa(y_{\star}^{(v)} ), \boldsymbol{\varphi}, \boldsymbol{\theta})$, and an approximated Markov Chain Monte Carlo (MCMC) sample from the posterior distribution
$\pi(\boldsymbol{\varphi}, \boldsymbol{\theta} \mid \mathcal{D})$. Based on the simulated sample, any quantity of interest associated with the predictive variables  can be easily extracted. In particular, predictions can be obtained for any node $Y^{(v)}$ conditional on a particular subset of observed nodes in the network. For instance, parent nodes could be predicted from their child nodes, or vice versa.

\subsection{Bayesian networks with mixed beta regression models} \label{subsec22}

Beta regression models offer a robust framework for analyzing continuous random variables strictly bounded within the interval $(0,1)$ in terms of a set of covariates and random effects. The beta probability density function can be parameterised using its mean $\mu \in (0,1)$ and a precision parameter $\phi > 0$ \citep{cribari2010beta}.

Consider a generic node defined by the random variable $Y_{ij}^{(v)}$, associated with the $j\text{-th}$ observation of the individual $i$, where $j=1,\dots,J$ and $i=1,\dots,n$. We assume conditional independence between the random variables $\boldsymbol{Y}^{(v)}=\{Y_{ij}^{(v)}, i=1,\dots, n, j=1,\dots, J\}$, given the vector of global parameters and hyperparameters, $\boldsymbol{\theta}$, and the vector of random effects, $\boldsymbol{\varphi}$, as follows:

\begin{equation*}
\begin{aligned}
    f(\boldsymbol{y}^{(v)} \mid Pa(\boldsymbol{y}^{(v)}),  \boldsymbol{\theta}, \boldsymbol{\varphi}) & = \prod_{i=1}^n \prod_{j=1}^J f({y}_{ij}^{(v)} \mid Pa(y_{ij}^{(v)}),  \boldsymbol{\theta}, \boldsymbol{\varphi}) \\
    & = \prod_{i=1}^n \prod_{j=1}^J \text{Be}(\mu_{ij}^{(v)}, \phi^{(v)}).
\end{aligned}
\end{equation*}

The conditional mean $\mu_{ij}^{(v)}$ of the beta distribution $\text{Be}(\mu_{ij}^{(v)}, \phi^{(v)})$ is linked to the parent nodes $Pa(Y_{ij}^{(v)})$ via a mixed regression structure with a suitable link function $g(\cdot)$. In this study, the logit is used as the link function as follows:

$$
g(\mu_{ij}^{(v)} \mid Pa(y_{ij}^{(v)}), \boldsymbol \theta^{(v)}, b_i^{(v)}) = \log \left( \frac{\mu_{ij}^{(v)}}{1 - \mu_{ij}^{(v)}} \right) = \beta_{0}^{(v)} +\sum_{u \in A_{ij}^{(v)}} \beta^{(v)}_{u} y_{ij}^{(u)} \,+\, b_i^{(v)},
$$
where $A_{ij}^{(v)} = \{ u \in \mathcal{V} \, : \, Y_{ij}^{(u)} \in Pa(Y_{ij}^{(v)})\}$. In this model,  the set of parameters and hyperparameters is $\boldsymbol{\theta^{(v)}} = (\boldsymbol{\beta}^{(v)}, \sigma_v^2)$, where $\boldsymbol{\beta}^{(v)}$ are the regression coefficients associated to the mixed regression model ($\beta_0^{(v)}, \beta_u^{(v)}, u \in A_{ij}^{(v)}$), and $b^{(v)}_i$ is the random effect associated with the individual $i$, assumed to be normally distributed as $(b^{(v)}_i \mid \sigma_v^2)\sim \text{N}(0, \sigma_{v}^2)$. It is worth mentioning that the set of random effects $\boldsymbol{b}^{(v)}=(b_1^{(v)}, \dots, b_n^{(v)})$ associated to all individuals of the sample are assumed identical and conditionally independent $(\boldsymbol{b}^{(v)} \mid \sigma_v) \sim \prod_{i=1}^n\text{N}(0,\sigma_v^2)$.

\section{Application to real data: Face expression vs emotions}\label{subsec3}

\subsection{Context of the study}\label{subsec31}

Computational modelling techniques are consolidating their role as essential pillars in the design, optimisation, and validation of advanced driver assistance systems. In particular, their role is especially relevant in the development of Connected and Autonomous Vehicles (CAVs) and in the conception of vehicles with higher levels of autonomy. These models, which range from vehicle dynamics to driver and pedestrian behaviour, are essential tools for defining the response of Advanced Driver Assistance Systems (ADAS) and for virtually validating their performance in complex scenarios. In this context, the ability to perform rapid inferences and generate immediate responses is critical, as decision-making must occur within fractions of a second to ensure system safety and efficiency \citep{hubmann2017decision}. 

In semi-autonomous driving settings, the interaction between human operators and vehicles is of paramount importance. A key aspect of this interaction lies in the interpretation of driver gestures and facial expressions, which provide relevant information about the driver’s level of attention, intention, and emotional state \citep{Egger2019}. However, the accurate modelling of these signals remains challenging due to the uncertain, dynamic, and noisy nature of the of the underlying variables. In this regard, probabilistic statistical models emerge as a robust methodology to capture and manage the inherent uncertainty of human behaviour. Since the pioneering work by \citep{forbes1995batmobile}, BN have demonstrated strong potential in state estimation, intention prediction, and decision-making under uncertainty, establishing themselves as suitable tools for modelling complex and dynamic systems such as automated driving. 

Despite the progress achieved since the mid-1990s, a widely validated and accepted model of human driving behaviour is still lacking \citep{abuali2016driver}. A reliable model is essential to enhance safety in both infrastructures and vehicles, particularly with the emergence of CCAM (Connected, Cooperative and Automated Mobility) technologies. Traditionally, three main modelling approaches have been considered: (i) deterministic models based on cognitive architectures, which are interpretable but limited in flexibility; (ii) artificial intelligence models, which can learn complex patterns but are often difficult to interpret and validate; and (iii) probabilistic models, which provide a more realistic representation by explicitly accounting for uncertainty and variability in both human and environmental conditions. 

In this context, Bayesian networks emerge as a strong candidate to address the current lack of validated driver behaviour models. Their ability to integrate heterogeneous information sources, manage uncertainty, and update beliefs in real time makes them particularly well suited for modelling dynamic human--vehicle interaction. Moreover, their interpretability provides a transparent alternative to black-box methods, facilitating validation and deployment within ADAS and CCAM systems.

\subsection{Dataset description} \label{subsec32}

The dataset utilized in this study is derived from an experiment conducted by the Institut de Biomecànica de València (IBV) within the framework of the European BERTHA project. The sample comprises 34 participants, including 18 men and 16 women.

The study is structured around two distinct scenarios: the first is characterized by a calm atmosphere designed to inhibit the expression of angry emotions, whereas the second features complex, rage-inducing situations. Throughout the scenarios, participants are repeatedly asked to report their subjective level of rage, $Y^{(R)}$. In addition, each scenario involves the monitoring of various facial variables within the Facial Action Coding System (FACS).

 FACS \citep{Ekman, Prince} is a well established methodology for recognizing and classifying facial expressions by coding specific facial muscle movements. Rather than assessing global expressions, FACS breaks them down into their smallest distinguishable movements, termed Action Units (AUs). Each AU produces a distinct change in facial appearance, such as raising an eyebrow or pursing the lips. In this study, the original dataset includes several continuous-valued variables strictly bounded within the $(0, 1)$ interval representing the activation levels of different facial muscle groups. These metrics capture the lifting of the outer eyebrow driven by the frontalis muscle, the frowning motion produced by the corrugator supercilii, the upper eyelid raising associated with the levator palpebrae superioris, and the cheek elevation linked to the orbicularis oculi. Additionally, they monitor the wrinkling of the nose area caused by the procerus and nasalis muscles, the upward movement of the mouth corners corresponding to the zygomaticus major, and the pressing of the lips against each other, which is typically associated with tension.
 
In the original experiment, each variable was monitored across five 60-second intervals per experimental phase, with five repeated measurements collected within each interval to capture the temporal evolution of facial and emotional responses. However, due to the low variability observed over time, these measurements were aggregated across temporal intervals for simplicity. A more detailed longitudinal representation would not yield additional relevant insights, but would unnecessarily increase model complexity. Consequently, a total of 68 aggregated records were obtained, evenly distributed with 34 records for each of the two experimental phases

\begin{figure} [H]
    \centering
    \includegraphics[width=0.98\textwidth]{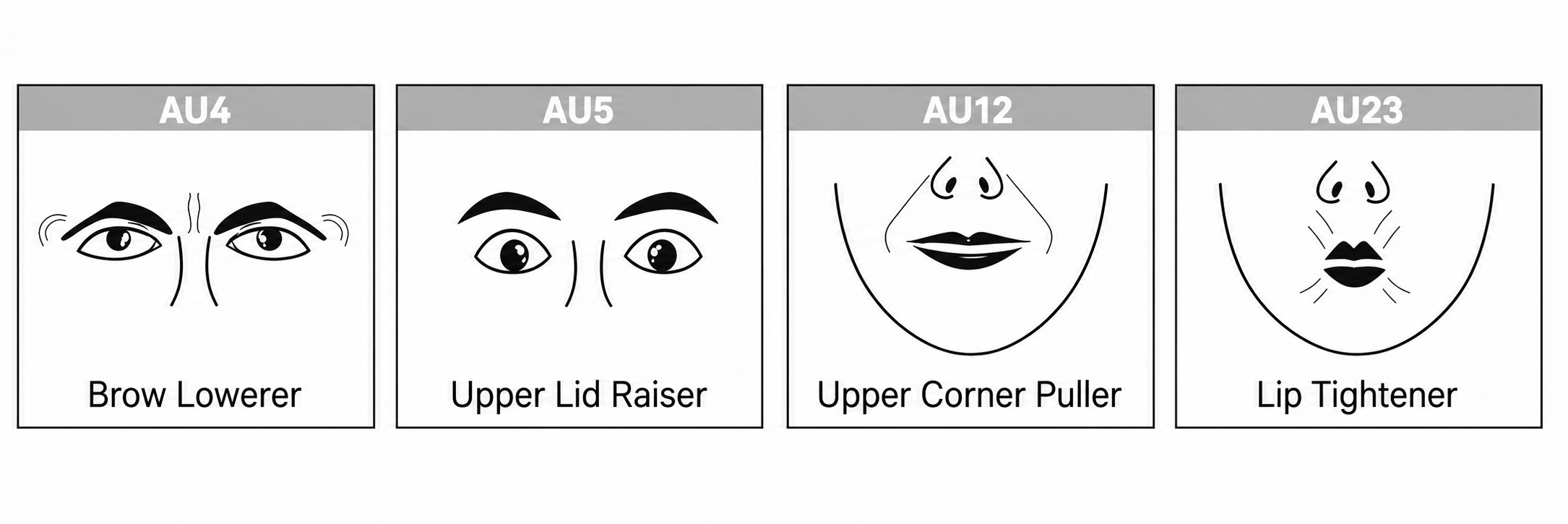}
\caption{Representative images of facial gesture patterns. The row shows, from left to right, the gestures corresponding to $Y^{(B)}$, $Y^{(E)}$, $Y^{(C)}$, and $Y^{(L)}$.
}
   \label{gestures}
\end{figure}

Exploratory data analysis revealed strong indications of multicollinearity among the initial facial gesture intensities. Consequently, a subset of these facial features was selected to model the expression of rage while minimizing redundant information. The final selection comprises the intensity of frowning, $Y^{(B)}$, upper eyelid raising, $Y^{(E)}$, lip corner pulling (smiling), $Y^{(C)}$, and lip pressing, $Y^{(L)}$ (see Figure~\ref{gestures}). 

\begin{figure} [H]
    \centering
    \includegraphics[width=0.85\textwidth]{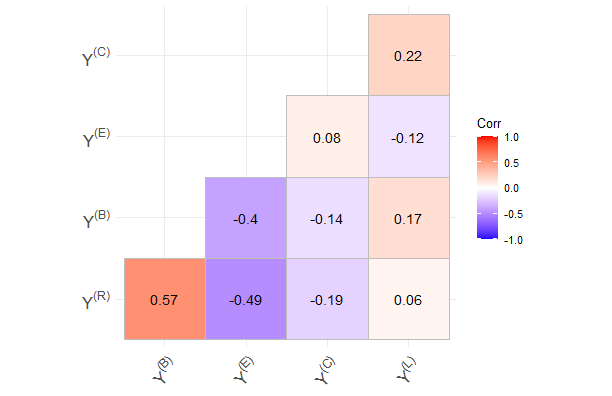}
\caption{Correlation matrix between facial gestures and rage level.
}
   \label{corr}
\end{figure}

Figure~\ref{corr} shows the correlation coefficient between the intensity of the selected facial gestures and the perceived degree of rage. The first thing that stands out  is  the moderate levels of association between each of the facial variables and the level of rage. Of particular note is the positive correlation of 0.57 between $Y^{(B)}$ and $Y^{(R)}$, and the negative correlation of -0.49 between $Y^{(E)}$ and $Y^{(R)}$. It is also important to note, as a very positive aspect, the weak linear relationship between the four facial variables. 

Descriptive statistics of the selected variables shows clear differences in the level and variability of the facial gesture variables, as shown in Table \ref{tabla}. In general, the variables present heterogeneous distributions, with differences in both central tendency and dispersion. In particular, $Y^{(B)}$ and $Y^{(L)}$ show the highest central values, with means of 0.4755 and 0.4805 respectively, and relatively high upper quartiles (0.6968 and 0.6196). This indicates that these gestures are frequently activated at moderate levels and are among the most present in the dataset. The variable $Y^{(E)}$ presents intermediate behaviour. Its mean (0.2934) and median (0.2823) are very similar, suggesting a stable distribution without strong skewness. Its values are also more concentrated compared to other gestures. In contrast, $Y^{(C)}$ shows consistently low values, with a mean of 0.0480 and a median of 0.0303. Its low quartiles (0.0212 for the first quartile) indicate that this gesture is rarely activated and appears mostly at low intensity. Finally, $Y^{(R)}$ shows the highest variability. It ranges from 0 to 1 and has a mean of 0.3809. The quartiles also show strong dispersion, with a median of 0.3000 and a third quartile of 0.7000, indicating that this gesture can vary from being completely inactive to highly activated depending on the observation.

\begin{table}[h]
\centering
\caption{Descriptive statistics of facial gesture variables}
\label{tabla}
\begin{tabular}{lccccc}
\hline
\textbf{Statistic} & \textbf{$Y^{(B)}$} & \textbf{$Y^{(E)}$} & \textbf{$Y^{(C)}$} & \textbf{$Y^{(L)}$} & \textbf{$Y^{(R)}$} \\
\hline
Minimum        & 0.1094 & 0.2126 & 0.0094 & 0.1122 & 0.0000 \\
1st Quartile   & 0.2561 & 0.2386 & 0.0212 & 0.3517 & 0.0000 \\
Median         & 0.4508 & 0.2823 & 0.0303 & 0.5034 & 0.3000 \\
Mean           & 0.4755 & 0.2934 & 0.0480 & 0.4805 & 0.3809 \\
3rd Quartile   & 0.6968 & 0.3235 & 0.0535 & 0.6196 & 0.7000 \\
Maximum        & 0.9121 & 0.5880 & 0.5573 & 0.7837 & 1.0000 \\
\hline
\end{tabular}
\end{table}

\subsection{Statistical modelling}\label{subsec33}

The experimental phase variable is a critical component of the modelling framework, accounting for both a baseline emotional state and a high-stress scenario designed to induce rage. Incorporating this variable is essential for capturing within-individual variability.

We evaluated two distinct modelling strategies for the phase variable: either treating it as an endogenous random variable within the network to account for its inherent uncertainty, or treating it as a fixed covariate. Although the latter approach is intuitive given the experimental control, modelling the phase as a random variable was preferred. This choice stems from the assumption that a driver's true latent state cannot be observed with absolute certainty, and that the spectrum of rage may encompass more than two distinct phases. Treating the phase as a random variable enables natural probabilistic characterization, facilitating marginal predictions of the rage level independent of the driver’s immediate state, as well as conditional predictions across other network nodes. Both approaches are rigorously compared regarding their inferential implications, predictive performance, interpretability, and overall statistical coherence.

 \subsubsection{Model with phase as a random variable}\label{subsec331}

Let $\boldsymbol{Y}^{(R)}=\{Y_{ij}^{(R)}, \.i=1,\ldots, n, \,j=0,1\}$ be the random vector that accounts for the random variables $Y_{ij}^{(R)}$ that measure the perceived degree of rage of individual $i$ at measurement $j$, where $j$ indexes the two measurements obtained for each individual. By analogy, we define the random vectors $\boldsymbol{Y}^{(B)}$, $\boldsymbol{Y}^{(E)}$, $\boldsymbol{Y}^{(C)}$, and $\boldsymbol{Y}^{(L)}$corresponding to the intensity of the facial gestures of each individual of the sample in each phase. Finally, let
$\mathbf{Y}^{(Ph)}=\left\{Y_{ij}^{(Ph)},\;i=1,\ldots,n,\;j=0,1\right\}$
denote the binary random vector indicating the phase associated with each
observation, where $Y_{ij}^{(Ph)}=0$ for the baseline phase and
$Y_{ij}^{(Ph)}=1$ for the stressful phase.

We define a BN for these random vectors, a vector of random effects $\boldsymbol{\varphi}$ and a vector $\boldsymbol \theta$ of parameters and hyperparameters as the joint probability distribution  
\begin{align}
f(\boldsymbol{y}^{(B)},  &\, \boldsymbol{y}^{(E)},  \boldsymbol{y}^{(C)}, \boldsymbol{y} ^{(L)},\boldsymbol{y}^{(R)}, \boldsymbol{y}^{(Ph)}, \boldsymbol{\varphi}, \boldsymbol{\theta}) = \nonumber \\
 & f(\boldsymbol{y}^{(B)},   \boldsymbol{y}^{(E)}, \boldsymbol{y}^{(C)}, \boldsymbol{y} ^{(L)},\boldsymbol{y}^{(R)}, \boldsymbol{y}^{(Ph)} \mid \boldsymbol{\varphi}, \boldsymbol{\theta})\, \,    f(\boldsymbol{\varphi} \mid  \boldsymbol{\theta}) \, \pi(\boldsymbol{\theta}),
\label{eq:eqva}
 \end{align}
\noindent where
\begin{align}
f(&\boldsymbol{y}^{(B)},  \boldsymbol{y}^{(E)}, \boldsymbol{y}^{(C)}, \boldsymbol{y} ^{(L)},\boldsymbol{y}^{(R)}, \boldsymbol{y}^{(Ph)} \mid \boldsymbol{\varphi}, \boldsymbol{\theta}) = \nonumber \\ &  \,f(\boldsymbol{y}^{(B)} \mid \boldsymbol{y}^{(R)}, \boldsymbol{y}^{(Ph)}, \boldsymbol{\theta})\, f(\boldsymbol{y}^{(E)} \mid \boldsymbol{y}^{(R)}, \boldsymbol{y}^{(Ph)}, \boldsymbol{\theta}) \, f(\boldsymbol{y}^{(C)} \mid \boldsymbol{y}^{(R)},\boldsymbol{y}^{(Ph)},  \boldsymbol{\theta}) \nonumber \\ &  \, f(\boldsymbol{y} ^{(L)} \mid \boldsymbol{y}^{(R)}, \boldsymbol{y}^{(Ph)}, \boldsymbol{\theta}) \,  f(\boldsymbol{y}^{(R)} \mid \boldsymbol{y}^{(Ph)}, \boldsymbol{\varphi}, \boldsymbol{\theta}) \, f(\boldsymbol{y}^{Ph} \mid \boldsymbol{\theta}) \nonumber \\ = & \,
\prod_{i=1}^{n}\prod_{j=0}^{1}\, f(y_{ij}^{(B)} \mid  y_{ij}^{(R)}, y_{ij}^{(Ph)}, \boldsymbol{\theta})\, f(y_{ij}^{(E)} \mid y_{ij}^{(R)}, y_{ij}^{(Ph)}, \boldsymbol{\theta}) \, f(y_{ij}^{(C)} \mid y_{ij}^{(R)}, y_{ij}^{(Ph)},  \boldsymbol{\theta}) \nonumber \\ &   \, f(y_{ij}^{(L)} \mid y_{ij}^{(R)}, y_{ij}^{(Ph)}, \boldsymbol{\theta}) \,  f(y_{ij}^{(R)} \mid y_{ij}^{(Ph)}, \boldsymbol{\varphi}, \boldsymbol{\theta}) \, f(y_{ij}^{Ph} \mid \boldsymbol{\theta}).
\label{eq:eqva}
\end{align} 
\noindent In this expression, $f(\boldsymbol \varphi \mid \boldsymbol \theta)$ is the conditional distribution of the random effects given $\boldsymbol \theta$, and $\pi(\boldsymbol \theta)$ is the prior distribution for $\boldsymbol \theta$.  Sex is included in the BN as a covariate to account for its potential influence on both rage levels and facial expressions.
\begin{figure} [H]
    \centering
    \includegraphics[width=0.8\textwidth]{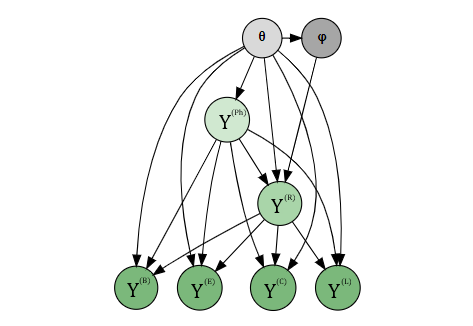}
    \caption{DAG of the proposed Bayesian network, where the phase variable is modelled as a random variable. The graph explicitly includes the random effects vector, as well as model parameters and hyperparameters, and their conditional dependencies.}
    \label{fig:bn1}
\end{figure}

Figure~\ref{fig:bn1} shows the DAG of the BN  in a general format in terms of the corresponding random vectors,  random effects, and  model parameters and hyperparameters. It is important to note that $\boldsymbol \theta$ is parent of all nodes,  random effects only have a direct effect on  rage level, and the phase is parent of all observable nodes.

We will now model the various conditional distributions that form part of the BN. 

 Phase $Y_{ij}^{(Ph)}$  is considered as a conditional  Bernoulli distribution 
\begin{equation*}
(Y_{ij}^{(Ph)} \mid  \boldsymbol{\theta} ) \sim \text{Ber}(p), \quad i = 1,\dots,34, \; j = 0,1,
\end{equation*}
where the parameter $p$ is common  to all individuals and phases.

The conditional distribution of the intensity $Y_{ij}^{(R)}$ of rage for individual $i$ at measurement $j$  is assumed to a mixed beta regression model.
\begin{equation*}
(Y_{ij}^{(R)} \mid y_{ij}^{(Ph)},  \boldsymbol{\theta}, \boldsymbol{\varphi}) \sim \text{Be}(\mu_{ij}^{(R)}, \phi^{(R)}), \quad i = 1,\dots,34, \; j = 0,1,
\end{equation*}
with the mean modelled using a logit link function and a linear predictor incorporating sex, phase, and a random intercept:
\begin{equation*}
\text{logit}(\mu_{ij}^{(R)}  \mid y_{ij}^{(Ph)},  \boldsymbol{\theta}, \boldsymbol{\varphi}) = \beta_0^{(R)} + \beta_M^{(R)} \, I_M(i) + \beta_{Ph}^{(R)} \, y_{ij}^{(Ph)} + b_i^{(R)}.
\end{equation*}
Here, $\beta_0^{(R)}$ is the intercept for women in the baseline phase, $\beta_M^{(R)}$ quantifies the effect of being a man, where $I_M(i)$ is an indicator variable taking the value 1 if individual $i$ is a man and 0 if woman, $\beta_{Ph}^{(R)}$ represents the effect of the stressful phase, and $b_i^{(R)}$ captures the individual-specific random intercept, accounting for the variability in $\mu_{ij}^{(R)}$ not explained by sex and the experimental phase, with $(b_i^{(R)}\mid \sigma^{(R)}) \sim \text{N}(0,\sigma^{(R)})$. 

Similarly, the intensity of the facial gesture variables $Y^{(G)}$ ($G \in \{B, C, E,  L\}$) is modelled conditionally on rage and phase:
\begin{equation*}
(Y_{ij}^{(G)} \mid y_{ij}^{(R)}, y_{ij}^{(Ph)},   \boldsymbol{\theta}) \sim \text{Be}(\mu_{ij}^{(G)}, \phi^{(G)}),
\end{equation*}
with the mean linked via the logit function:
\begin{equation*}
\text{logit}(\mu_{ij}^{(G)}\mid y_{ij}^{(R)}, y_{ij}^{(Ph)},   \boldsymbol{\theta}) = \beta_0^{(G)} + \beta_M^{(G)} \, I_M(i) + \beta_{Ph}^{(G)} \, y_{ij}^{(Ph)} + \beta_{R}^{(G)} \, y_{ij}^{(R)}.
\end{equation*}
As above, $\beta_0^{(G)}$ is the intercept for women, $\beta_M^{(G)}$ captures the effect of being men, where $I_M(i)$ is a sex indicator variable, $\beta_{Ph}^{(G)}$ represents the effect of the experimental phase (provocation versus baseline), and $\beta_{R}^{(G)}$ quantifies the influence of rage on the gesture, accounting for sex and phase.

The Bayesian model is completed by specifying a prior distribution  for all model parameters and hyperparameters. We assume prior independence among them and a scenario with very little prior information.  We elicit a normal distribution  $\text{N}(0, 10^2)$   to all regression coefficients, including intercepts and slopes. This choice reflects the absence of strong prior assumptions while allowing sufficient flexibility for parameters governing responses bounded in $[0,1]$. For the   dispersion parameter of the beta distributions, a gamma distribution, $\text{Ga}(1, 0.1)$, is assigned; a   non-informative beta distribution, Be(1,1) is considered for the parameter $p$ of the Bernoulli distribution associated with the phase, and a  wide uniform distribution, $\text{U}(0,1000)$, is applied to the hyper-prior standard deviations governing the random effects. 

\subsubsection{Model with phase as a  covariate}\label{subsec332}

\begin{figure} [H]
    \centering
    \includegraphics[width=0.8\textwidth]{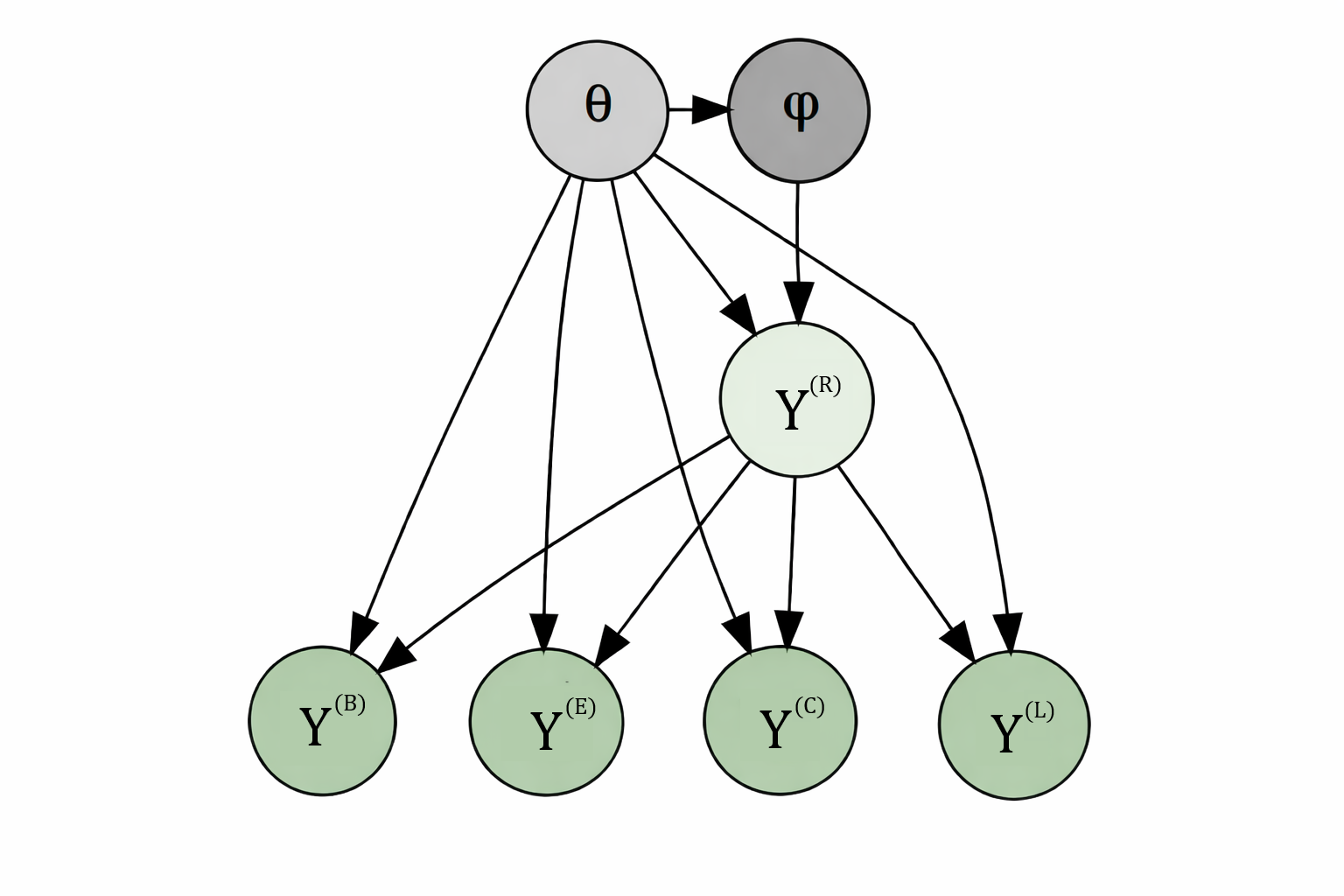}
    \caption{DAG of the proposed BN, where the phase variable is modelled as a fixed covariate. The graph explicitly includes the random effects vector, as well as model parameters and hyperparameters, and their conditional dependencies.}
    \label{fig:bn2}
\end{figure}

 This alternative model mirrors the previous configuration, with the sole distinction that the phase is treated as a covariate; consequently, it lacks a probability distribution and is excluded as a node from the network. Retaining the notation established for the first model, the associated BN, whose structure is illustrated in Figure \ref{fig:bn2}, is defined by the following joint probability distribution:
\begin{align}
f(\boldsymbol{y}^{(B)},   \boldsymbol{y}^{(E)}, & \boldsymbol{y}^{(C)}, \boldsymbol{y} ^{(L)},\boldsymbol{y}^{(R)}, \boldsymbol{\varphi}, \boldsymbol{\theta}) = \nonumber\\
 & \, f(\boldsymbol{y}^{(B)},   \boldsymbol{y}^{(E)}, \boldsymbol{y}^{(C)}, \boldsymbol{y} ^{(L)},\boldsymbol{y}^{(R)} \mid \boldsymbol{\varphi}, \boldsymbol{\theta})\, \,    f(\boldsymbol{\varphi} \mid  \boldsymbol{\theta}) \, \pi(\boldsymbol{\theta}),
\label{eq:eqcv}
 \end{align}

\noindent where 
\begin{align}
f(\boldsymbol{y}^{(B)},&  \boldsymbol{y}^{(E)}, \boldsymbol{y}^{(C)}, \boldsymbol{y} ^{(L)},\boldsymbol{y}^{(R)}  \mid \boldsymbol{\varphi}, \boldsymbol{\theta}) = \nonumber \\   & \,f(\boldsymbol{y}^{(B)} \mid \boldsymbol{y}^{(R)},   \boldsymbol{\theta})\, f(\boldsymbol{y}^{(E)} \mid \boldsymbol{y}^{(R)},  \boldsymbol{\theta}) \, f(\boldsymbol{y}^{(C)} \mid \boldsymbol{y}^{(R)},  \boldsymbol{\theta})     \nonumber \\ &  f(\boldsymbol{y} ^{(L)} \mid \boldsymbol{y}^{(R)}, \boldsymbol{\theta}) \, f(\boldsymbol{y}^{(R)} \mid \boldsymbol{\varphi}, \boldsymbol{\theta}) \, \nonumber \\ = & \,
\prod_{i=1}^{n}\prod_{j=0}^{1}\, f(y_{ij}^{(B)} \mid  y_{ij}^{(R)},  \boldsymbol{\theta})\, f(y_{ij}^{(E)} \mid y_{ij}^{(R)}, \boldsymbol{\theta}) \, f(y_{ij}^{(C)} \mid y_{ij}^{(R)},  \boldsymbol{\theta}) \nonumber \\ & f(y_{ij}^{(L)} \mid y_{ij}^{(R)}, \boldsymbol{\theta}) \, f(y_{ij}^{(R)} \mid \boldsymbol{\varphi}, \boldsymbol{\theta}),  
\label{eq:eqva}
\end{align} 

\noindent $f(\boldsymbol \varphi \mid \boldsymbol \theta)$ is the conditional distribution of the random effects given $\boldsymbol \theta$, and $\pi(\boldsymbol \theta)$ is the prior distribution for $\boldsymbol \theta$.  Sex is also included in the BN as a covariate.

Figure~\ref{fig:bn1} shows the DAG of the BN  in a general format in terms of the random vectors for each of the observed vectors, the random effects, and the model parameters and hyperparameters.

As previously noted, the phase variable in this model is deterministic, serving strictly as a covariate. Because the stochastic behavior of the rage level and facial gesture intensities is characterized via conditional distributions, their mathematical formulation remains completely analogous to that of the random-phase model. This structural equivalence extends to the conditional distribution of the random effects given the model parameters and hyperparameters, as well as to their respective prior distributions.

\section{Posterior outputs}\label{sec4}

The posterior distribution associated with the quantities of each of the two models has been implemented via Markov chain Monte Carlo methods  with 
WinBUGS software \citep{lunn2000winbugs}.
Computation was carried out  running three parallel chains for each
model with a total of 4,000,000 iterations and a burn-in of
500,000. To reduce autocorrelation in the
sample, we also thinned the chains by storing every 1,300th iteration.

All analyses were implemented in R. The code supporting the study is available at \url{https://github.com/zairamndez13/Bayesian-networks-gesture-driving}. Data cannot be shared due to confidentiality constraints.

\subsection{Estimation}\label{subsec41}

From an inferential standpoint, modelling phase as either a random effect or a fixed covariate leads to essentially equivalent estimation results. Posterior summaries for both specifications are presented in Table \ref{fig:comparison_models}. For brevity, the numerical results discussed in the text refer to the model with phase as a random variable, as differences between specifications are negligible.
 
Posterior inferences for the regression coefficients associated with the rage mean level $\mu^{(R)}$ show that the sex effect parameter $\beta_M^{(R)}$ has a posterior mean close to zero and a 95\% credible interval of $[-0.59, 0.97]$. This suggests that there is no evidence of differences between men and women. In contrast, the phase effect parameter $\beta_{Ph}^{(R)}$ exhibits a strong positive posterior shift, with a 95\% credible interval of $[2.70, 4.25]$, clearly excluding zero and indicating a substantial increase in rage during the stressful phase. This information is presented graphically in Figure \ref{fig:comparison_models}, which shows the posterior distribution of the mean rage by sex and phase. The figure highlights that the experimental phase is the dominant source of variation in  the level of rage, whereas sex differences are negligible due to the substantial overlap in the  posterior distribution of both  groups.

\begin{table}[H]
\centering
\begin{tabular}{lcccc|cccc}
\hline
& \multicolumn{4}{c}{Phase random} 
& \multicolumn{4}{c}{Phase covariate} \\
\cline{2-5} \cline{6-9}
Parameter & Mean & Sd & 2.5\% & 97.5\% & Mean & Sd & 2.5\% & 97.5\% \\
\hline
$\beta_0^{(R)}$ & -2.58 & 0.38 & -3.39 & -1.86 & -2.54 & 0.39 & -3.32 & -1.82 \\
$\beta_M^{(R)}$ & 0.18 & 0.39 & -0.59 & 0.97 & 0.14 & 0.37 & -0.56 & 0.90 \\
$\beta_{Ph}^{(R)}$ & 3.50 & 0.40 & 2.70 & 4.25 & 3.45 & 0.39 & 2.66 & 4.19 \\
$\sigma^{(R)}$ & 0.77 & 0.31 & 0.11 & 1.37 & 0.71 & 0.32 & 0.07 & 1.30 \\
\midrule
$\beta_0^{(B)}$ & -0.91 & 0.18 & -1.26 & -0.56 & -0.91 & 0.18 & -1.27 & -0.56 \\
$\beta_M^{(B)}$ & 0.41 & 0.19 & 0.02 & 0.79 & 0.41 & 0.19 & 0.03 & 0.78 \\
$\beta_{Ph}^{(B)}$ & 0.36 & 0.32 & -0.26 & 0.99 & 0.36 & 0.32 & -0.27 & 0.98 \\
$\beta_{R}^{(B)}$ & 1.10 & 0.44 & 0.21 & 1.95 & 1.10 & 0.44 & 0.21 & 1.96 \\
\midrule

$\beta_0^{(E)}$ & -0.66 & 0.06 & -0.78 & -0.53 & -0.66 & 0.06 & -0.78 & -0.53 \\
$\beta_M^{(E)}$ & -0.03 & 0.07 & -0.18 & 0.11 & -0.04 & 0.07 & -0.17 & 0.11 \\
$\beta_{Ph}^{(E)}$ & -0.43 & 0.12 & -0.68 & -0.19 & -0.44 & 0.12 & -0.68 & -0.20 \\
$\beta_{R}^{(E)}$ & 0.03 & 0.17 & -0.32 & 0.36 & 0.03 & 0.17 & -0.30 & 0.35 \\
\midrule

$\beta_0^{(L)}$ & -0.28 & 0.15 & -0.57 & 0.00 & -0.29 & 0.15 & -0.59 & 0.00 \\
$\beta_M^{(L)}$ & 0.25 & 0.16 & -0.07 & 0.56 & 0.25 & 0.16 & -0.07 & 0.57 \\
$\beta_{Ph}^{(L)}$ & 0.30 & 0.26 & -0.22 & 0.83 & 0.29 & 0.27 & -0.22 & 0.83 \\
$\beta_{R}^{(L)}$ & -0.22 & 0.36 & -0.93 & 0.49 & -0.19 & 0.36 & -0.89 & 0.52 \\
\midrule

$\beta_0^{(C)}$ & -2.90 & 0.17 & -3.25 & -2.57 & -2.90 & 0.17 & -3.24 & -2.56 \\
$\beta_M^{(C)}$ & 0.22 & 0.18 & -0.14 & 0.58 & 0.22 & 0.19 & -0.14 & 0.58 \\
$\beta_{Ph}^{(C)}$ & -0.21 & 0.28 & -0.76 & 0.34 & -0.20 & 0.28 & -0.76 & 0.35 \\
$\beta_{R}^{(C)}$ & -0.09 & 0.40 & -0.88 & 0.67 & -0.11 & 0.40 & -0.89 & 0.67 \\
\hline
\end{tabular}
\caption{Posterior summaries for both models: phase as a fixed covariate and as a random effect.}
\label{fig:comparison_models}
\end{table}
 
Regarding facial expressions, brow lowering  shows low baseline activation in women during the baseline phase. The intercept parameter $\beta_0^{(B)}$ has a negative posterior mean, with a 95\% credible interval of $[-1.26,-0.56]$, indicating a reduced baseline probability of activation.

\begin{figure} [H]
    \centering
    \includegraphics[width=0.9\textwidth]{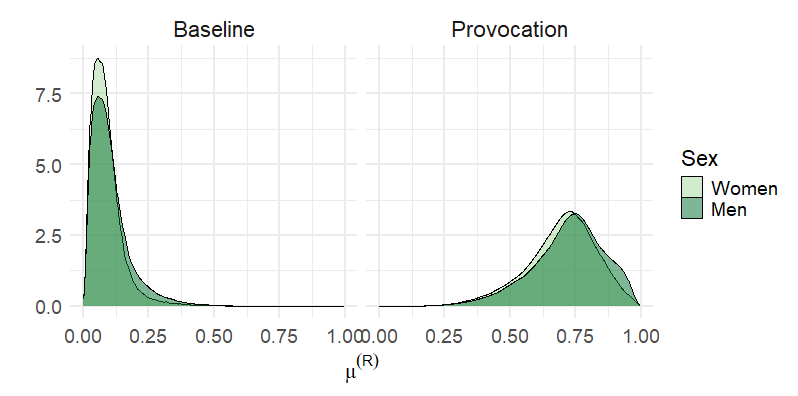}
    \caption{Posterior distribution of the mean rage level according to the individual’s sex and the experimental phase.}
    \label{fig:dp_vafase}
\end{figure}

The posterior mean of the regression coefficient associated with being a man as compared to being a woman is 0.41  with a 95\% credible interval of $[0.02,0.79]$. Therefore, there is evidence that men have a higher probability of exhibiting brow lowering compared to women holding other variables constant. 
The posterior mean associated with the stressful phase is 0.36 with a 95\% credible interval   $[-0.26,0.99]$, suggesting that there is no relevant effect of the phase on gesture activation. The regression coefficient associated with rage   shows a clear positive  effect  indicating that higher rage levels are associated with an increased probability of activating brow lowering. 

The estimated  model for upper lid raising  shows moderate baseline activation in women during the baseline phase. The intercept parameter $\beta_0^{(E)}$ has a negative posterior mean, with a 95\% credible interval of $[-0.78,-0.53]$, suggesting a lower baseline probability of activation. The parameter associated with the difference in impact between men and women does not provide evidence of a meaningful difference between them. The posterior information on the regression coefficient for the phase   shows a negative posterior effect with a 95\%  credible interval of $[-0.68,-0.19]$, indicating a reduction in the probability of activation during provocation. On the other hand, rage does not seem to play a relevant role in the activation of this expression.

The posterior summaries  for lip tightener ($Y^{(L)}$) suggest that neither baseline activation nor any of the covariates, phase, level of rage and gender,   show a meaningful association with the intensity of  this expression. For upper corner puller ($Y^{(C)}$), the intercept parameter $\beta_0^{(C)}$ shows a clear posterior effect, with a 95\% credible interval of $[-3.25,-2.57]$, indicating a reduced baseline probability of activation. The regression coefficients associated with gender, phase, and rage all have credible intervals that include zero, providing no evidence of their effects on expression activation.

In summary, brow lowering is the expression with the strongest positive association with rage intensity, more frequently expressed by men and little affected by the experimental phase. Upper lid raising displays a distinct pattern, characterised by a relevant decrease in activation during the provocation phase and no clear association with rage or sex. This suggests that this expression may be more sensitive to contextual or attentional factors than to emotional intensity. In contrast, lip tightener and the upper corner puller are characterised by higher posterior uncertainty and low association with   
rage, sex, and phase. Overall, the results point to a strong contextual modulation of rage combined with gesture-specific expressive dynamics. Although rage intensity responds robustly to provocation, only selected facial gestures, most notably brow lowering, exhibit a
clear and systematic relationship with emotional intensity.

\subsection{Prediction}\label{subsec42}

One of the main objectives of this  study is to develop a BN that is capable of predicting the intensity of rage in people with specific levels of facial expressions.
This predictive ability is particularly relevant in applied contexts, as it enables
the anticipation of information about the degree of emotional states potentially associated with risky driving
behavior, thereby providing a methodological foundation for intelligent driver
assistance systems.

To assess the predictive performance of both modelling strategies (phase as a random variable or as a covariate), we use the Watanabe-Akaike Information Criterion (WAIC). This criterion provides an estimate of out-of-sample predictive precision by integrating across the entire posterior distribution \citep{gelman2014understanding}. Furthermore, WAIC is asymptotically equivalent to leave-one-out cross-validation and generally provides a more stable and robust assessment in hierarchical and non-regular models. WAIC results have been  computed using the \texttt{loo} package in R \citep{vehtari2017practical}. 

\begin{table}[H]
\centering
\begin{tabular}{lc}
\hline
Model & WAIC \\
\hline
Phase random  & -802.6 \\
Phase covariate & -802.6 
\\
\hline
\end{tabular}
\caption{Model comparison using the WAIC.}
\label{tab:dic_waic}
\end{table}

Table \ref{tab:dic_waic} reports the corresponding WAIC values. Since both models exhibit identical predictive performance, we select the formulation in which phase is treated as a random variable, as it allows predictions  to be performed regardless of whether or not the driver’s current state is known at the time the prediction is made (for example, relaxed or angry).

The prediction of new experimental results is based on sampling from the joint posterior
predictive distribution of the observable nodes of the BN,
\begin{align}
f(&y^{(Ph)}_{\star},   y^{(R)}_{\star}, y^{(B)}_{\star}, y^{(C)}_{\star}, y^{(L)}_{\star}, y^{(E)}_{\star} \mid \mathcal{D})= \nonumber\\
&\int f(y^{(Ph)}_{\star}, y^{(R)}_{\star}, y^{(B)}_{\star}, y^{(C)}_{\star}, y^{(L)}_{\star}, y^{(E)}_{\star} \mid \boldsymbol{\varphi}, \boldsymbol{\theta})\, \pi(\boldsymbol{\varphi},\boldsymbol{\theta} \mid \mathcal{D})\, \text{d}(\boldsymbol{\varphi},\boldsymbol{\theta}),
\label{eqn:predi}
\end{align}
 
\noindent where $f(y^{(Ph)}_{\star}, y^{(R)}_{\star}, y^{(B)}_{\star}, y^{(C)}_{\star}, y^{(L)}_{\star}, y^{(E)}_{\star} \mid \boldsymbol{\varphi}, \boldsymbol{\theta})$ is the same conditional distribution as that in (\ref{eq:eqva}). The joint predictive distribution allows us to approximate any marginal and conditional posterior predictive distribution. In our case, our focus will be on obtaining posterior predictions of the level of rage based on the observed intensities of facial expressions, specifically brow lowering   and upper
eyelid raising  because they   showed the strongest inferential relevance in the quantification of  the degree of rage.  

\begin{figure}[H]
  \centering
  \begin{subfigure}[b]{0.49\textwidth}
    \centering
    \includegraphics[width=\textwidth]{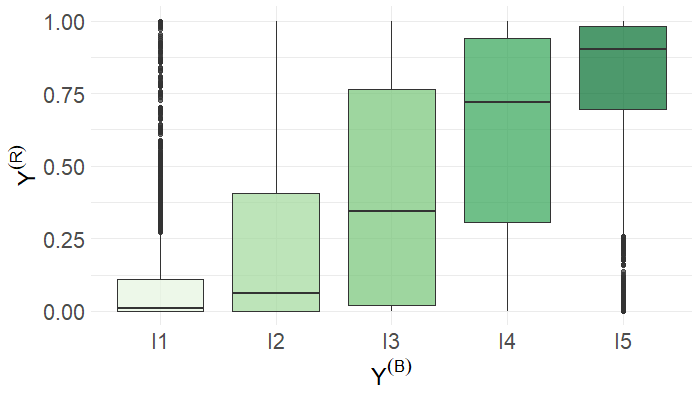}
    \label{fig:predictiva-mujer}
  \end{subfigure}
  \hfill
  \begin{subfigure}[b]{0.48\textwidth}
    \centering
    \includegraphics[width=\textwidth]{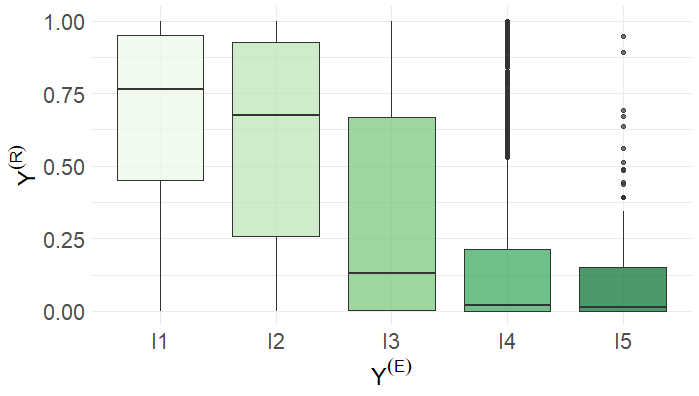}
    \label{fig:predictiva-hombre}
  \end{subfigure}
  \caption{Posterior predictive distribution of the intensity of the rage, $Y^{(R)}$,  conditional on the different categories of the intensity of the  brow lowering (left-hand graph) and upper eyelid lifting (right-hand graph). In both figures, category 1 corresponds to very low intensity, 2 to low,    3 to moderate, 4 to strong and 5 to very strong. }
  \label{fig:predictiva-va-mosaico}
\end{figure}

Graphs with many variables are not very easy to understand. With this in mind and to make the results easier to interpret, we have decided to discretise the variables that capture the information on facial gestures. We have defined a  symmetric and balanced Likert scale  composed of five levels for measuring the a very weak (1), weak (2), moderate (3), strong (4) and very strong (5) activation of each facial gesture.   
The two facial expressions most closely associated with rage levels are brow lowering and upper lid raising. In this regard, Figure \ref{fig:predictiva-va-mosaico} shows the posterior predictive distribution of the level of rage according to  the different intensity categories for brow lowering (left panel) and upper lid raising (right panel), 
$f(y_{\star}^{(R)} \mid y_{\star}^{(B)}, \mathcal D)$ and 
$f(y_{\star}^{(R)} \mid y_{\star}^{(E)}, \mathcal D)$, respectively.  \\

A clear monotonic relationship is observed between brow lowering and rage intensity: the level of rage will increase progressively across higher activation of brow lowering, indicating that stronger activation of this expression is associated with higher predicted levels of rage. In contrast, the posterior predictive distribution of the intensity of the rage  conditional on the level of upper lid raising exhibits an inverse pattern. Higher predicted rage levels are concentrated in the lower activation intervals of upper eyelid lifting, whereas the posterior predictive medians decrease as activation increases. This negative association suggests that upper lid raising may serve a distinct expressive or regulatory function, potentially reflecting mechanisms of emotional containment rather than emotional escalation. 

Predictive analysis of the level of rage in relation to each expression individually does not capture the full information about this emotion contained within the network.  To harness the full potential of prediction in (\ref{eqn:predi}) to obtain information about  the level of rage in relation to all the facial expressions considered, we   used clustering techniques to identify patterns of individuals with different facial expression profiles. Unlike approaches based on
predefined thresholds, clustering partitions the observations according to the structure present in the data, allowing facial expression
profiles to emerge naturally from the predicted gesture intensities. The resulting groups define three distinct driver profiles, ranging from low to high levels of emotional activation. Consequently, we labeled these groups as A, B, and C, based on the intensity of their facial expressions. 

Table \ref{tab:profiles}  presents the primary characteristics, specifically the mean along with the first ($Q_1$) and third ($Q_3$) quartiles of intensit, for the four facial expressions, across the three identified individual profiles.   Profile C   is
characterised by elevated levels of brow lowering and lip tightening
together with lower levels of upper eyelid raising; conversely, Profile A exhibits an opposite pattern. 
Profile B occupies an intermediate position between these two extremes. Lip corner pulling (smiling)  exhibits a less pronounced pattern, attaining its highest levels in the B group while remaining comparatively low in both the A and C profiles. Overall, these trends are consistent with the relationships previously identified through the inferential and predictive analyses. 

\begin{table}[H]
\centering

\begin{tabular}{ccccc}
\hline
Profile  &  $Y^{(B)}$ & $Y^{(E)}$ & $Y^{(C)}$ &  
$Y^{(L)}$\\
\hline

A &
0.31 (0.18, 0.42) &
0.44 (0.32, 0.55) &
0.04 (0.02, 0.06) &
0.34 (0.30, 0.38)
 \\

B &
0.47 (0.30, 0.62) &
0.48 (0.35, 0.60) &
0.13 (0.10, 0.15)  &
0.30 (0.25, 0.35)\\

C &
0.65 (0.52, 0.79) &
0.52 (0.40, 0.64) &
0.03 (0.01, 0.05)  &
0.24 (0.21, 0.28)\\

\hline
\end{tabular}
\caption{Summary of the facial-expression characteristics associated with the identified driver profiles. Values are reported as mean and quartiles (Q1,   Q3).}
\label{tab:profiles}
\end{table}

\begin{figure} [H]
    \centering
    \includegraphics[width=0.9\textwidth]{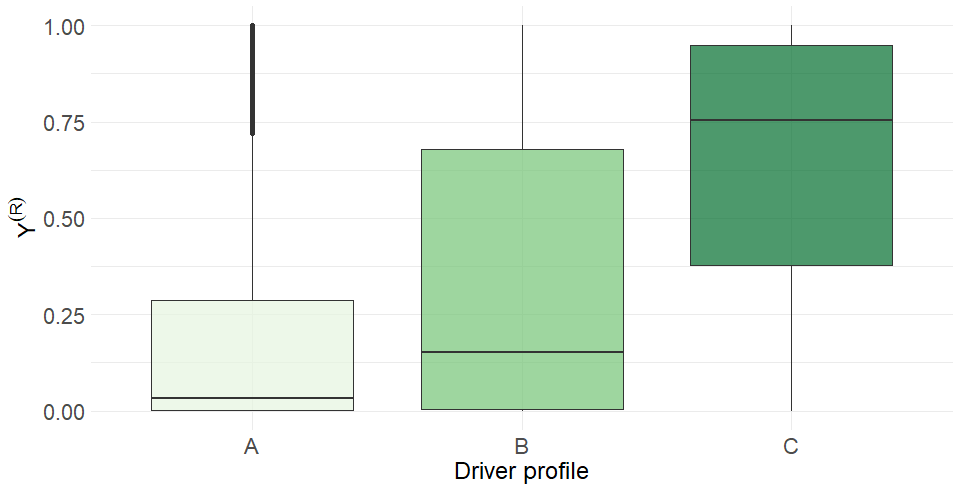}
    \caption{Distribution of rage intensity, $Y^{(R)}$, across the driver profiles identified through clustering of the predicted facial expressions.}
    \label{fig:d_profile}
\end{figure}

Figure \ref{fig:d_profile} displays the posterior predictive distribution of the  level of rages with regard to facial profiles A, B and C. Despite the wide variability in the results, important differences can be observed in the predicted distribution of rage across the different individual profiles. In the case of Profile A, we have fairly low  predictions of the level of rabies (mean=0.20 and quartiles Q1=0.00 and Q3=0.29), although some outliers are also observed;  the predicted level of rage in individuals with profile B is slightly higher (mean=0.34 and quartiles Q1=0.00 and Q3=0.68). This difference is considerably greater in the case of individuals with profile C: the mean is 0.64 with quartiles Q1=0.38 and Q3=0.95.

\section{Conclusions}\label{sec5}

This article combines Bayesian networks (BNs), simulation-based inference and mixed beta regression models from a fully Bayesian statistical perspective. It  provides an effective framework for modelling the relationship between specific facial expressions and emotions in human driving environments. 

From a modelling point of view, a key issue has been how to treat the experimental variable ‘phase’. To address this, two alternative formulations were considered: one in which phase was included as a fixed covariate, and another in which it was modelled as a random variable in the BN.  It is important to note that in both cases, phase plays a particularly significant role in modelling rage intensity.  In our study, formulating  phase as a random variable offers a clear practical advantage for prediction. Specifically, for any posterior predictive distribution of interest for rage, it is possible to integrate out the uncertainty regarding phase, which is impossible when phase is treated as a covariate. This is particularly relevant in real-world driving contexts, where the stressful phase is not directly observable.

Brow lowering is identified as the expression with the strongest positive association with rage, showing greater activation in men than in women. In contrast, upper lid raising exhibited an inverse relationship: higher activation predicted lower rage levels, indicating that its expression tends to diminish as emotional intensity increases. Other expressions, such as lip tightening and lip corner pulling, exhibited less clear relationships with rage.

The model’s ability to predict rage on the basis of facial expressions constitutes the most important output of this paper within the context of human driving. Because the uncertainty associated with prediction is inherently greater than at the estimation stage, it is desirable to incorporate more experimental data or, failing that, expert-validated information. A more complex experimental design involving additional nodes or demographic and social covariates could also provide deeper insights into the relationship between facial expressions and emotions

Overall, these findings illustrate the strength of BN in both inferential and predictive tasks. Their capacity to integrate heterogeneous information, manage uncertainty, and update beliefs in real time makes them highly suitable for modelling dynamic human–vehicle interactions. Furthermore, their interpretability provides transparency that is often lacking in black-box AI methods, facilitating validation and deployment in real-world Advanced Driver Assistance Systems (ADAS) and semi-autonomous driving scenarios. 

\section*{Acknowledgments}\label{sec6}
Research activity under BERTHA project (GA101076360) funded by the European Union. Views and opinions expressed are however those of the authorC only and do not necessarily reflect those of the European Union or the European Climate, Infrastructure and Environment Executive Agency (CINEA). Neither the European Union nor the granting authority can be held responsible for them.

This work is part of the project PID2022-136455NB-I00, funded by Ministerio de Ciencia, Innovación y Universidades of Spain (MCIN/AEI/10.13039/501100011033/ FEDER, UE) and the European Regional Development Fund.

\bibliographystyle{chicago}

\end{document}